\documentclass[oneside,namedreferences,english]{kluwer}

\usepackage{epsfig}

\usepackage[T1]{fontenc}
\usepackage[latin1]{inputenc}
\usepackage{babel}
\usepackage{setspace}

\begin{document}





\thispagestyle{empty}

\hbox{}

\centerline{\Large \bf The effect of increasing the rate of repetitions  }

\centerline{\Large \bf  of classical reactions } 

\bigskip  \bigskip \bigskip \bigskip \bigskip \bigskip \bigskip \bigskip

\centerline{\Large \bf D. Bar \footnote{Department of Physics, 
Bar-Ilan University, 
Ramat-Gan, Israel}}

\bigskip     \bigskip     \bigskip    \bigskip \bigskip

\bigskip    \bigskip \bigskip \bigskip \bigskip \bigskip

\underline{\bf e-mail}:\ \ \ \ \ \ \ \ \ \ \ \ \ \ \ \ \ \ \ \ \ \ \ \ \ \ \ \ \
{\bf \large  bardan@mail.biu.ac.il}

\bigskip    \bigskip


\begin{article}
\begin{opening}
\title{The effect of increasing the rate  of
repetitions of classical reactions }

\makeatother
\bigskip \bigskip
 
\author{ D.Bar } 
\institute{Department of Physics, Bar Ilan University, Ramat Gan,
Israel }
\runningtitle{The  effect of increasing the rate.....}

 \runningauthor{ Bar}
\bigskip
 
\begin{abstract}   \noindent 
{\it Using quantum theory operator methods  we  discuss the general reversible reactions 
$A_1+A_2+\cdots A_r \leftrightarrow B_1+B_2+\cdots +B_s$, 
where $r$ and $s$  are arbitrary natural 
positive numbers.    We
show that if  either direction of the  reaction is  repeated a large number 
of times  $N$   
 in a finite total
time $T$  then in the limit of very large $N$, keeping $T$ constant,  
one remains with the initial 
reacting particles  only. We also show that if the 
reaction 
evolves through different possible paths of evolution, each of them beginning at
the same side of the reaction, proceeds through different intermediate 
consecutive 
reactions and ends at the other side, then  one may  ``realize''  any 
such path by
performing in a dense manner   the  set of reactions along it.  The same results
are  also 
  numerically  demonstrated 
  for   the specific reversible reaction $A+B 
  \leftrightarrow A+C$. We note that  similar  results have been shown to hold 
  also   in 
the quantum regime. }
   
\end{abstract}
\noindent


\noindent 
\bigskip 

\end{opening}

\bigskip

\noindent 
\protect \section{Introduction \label{sec1}}
\noindent The general reversible reactions  
 $A_1+A_2+\cdots+ A_r\leftrightarrow B_1+B_2+\cdots +B_s$, where 
$r$ and $s$  are two arbitrary positive 
natural numbers,  have
been studied by many authors (see, for example, \cite{Havlin00} and references
therein). These studies discuss, 
especially,  the effects of the single reaction, or, in case it is repeated $N$
times,  the effect of these repetitions where
the general total time $T$  increases proportionally to $N$. We
can, however, imagine a situation in which the {\it rate} of these repetitions
increases and discuss the effect of this increase upon the reaction. Such an
effect have been studied in \cite{Kac47} with respect to random walk and it was
shown that when the rate of repeating this  walk becomes very large one
obtains a Brownian motion \cite{Nelson67}. It has also been shown 
\cite{Bar01}, \cite{Bar03} 
 with respect to 
the finite 
one-dimensional  array of $N$ imperfect traps \cite{Havlin00}, \cite{Smol17} 
 that as $N$ 
  becomes very large   the survival probability 
\cite{Havlin00} of the 
particles that pass through them tends to unity. We show in this work that if 
 either direction of the  mentioned  reversible reaction is repeated  a
large number  $N$ of times  in a finite total time $T$, then in the limit 
of very
large $N$  (where  $T$ is kept constant)  one remains with the  initial 
reacting 
particles only.  \par We  use   quantum theory methods and   terminology  
      as done by  
many authors that use quantum formalism  for analysing
classical reactions (see for example \cite{Mattis98} and annotated bibligraphy
therein).  The most suitable quantum method is the coherent state one
\cite{Glauber63},  \cite{Klauder68} which  
  allows us  to define simultaneously, 
  as has been remarked in \cite{Swanson92}, 
the expectation values of the  
 conjugate variables $Q$ and $P$.   Thus,   they   may both  
 have nonzero
values in which case the entailed  formalism resembles \cite{Swanson92} 
the classical one and so it may be 
  used for discussing   classical reactions. \par   The use of the coherent
state formalism, together with  second quantization methods, for classical
systems have been extensively studied by Masao \cite{Masao76}.    
   Since the
described phenomena and,   especially,  the particles participating 
in the reactions 
 are classical we  represent them  here 
 by {\it real } 
coherent states. 
That is,   we  denote 
    the reacting  and  product particles 
by  the real coherent states   
$|z\!>= e^{-\frac{1}{2}|z|^2}\sum_{n=0}^{n=\infty}\frac{z^n}{(n!)^{\frac{1}{2}}}|n\!>$,
where $z$ is the real number 
$z=\frac{q+p}{(2)^{\frac{1}{2}}}$, $q$ and $p$ are two arbitrary real $c$
numbers  and $|n\!>$ are number representation eigenstates  
\cite{Klauder68}. The masses of the reacting and product particles are assigned, for 
convenience,  the unity value. 
We note that although the mathematical 
entities and "operators" 
involved in this method  do not conform, as will be shown, to the known 
quantum operator formalism,  we, nevertheless,   follow, except for the differences,  the 
 conventional definitions and methods of the last theory. 
 The results  obtained 
by applying the   
real coherent state formalism for classical reactions 
are exactly the same as those previously obtained \cite{Bar00} 
by applying the complex coherent state methods for 
 quantum reactions. 
Moreover, although the real coherent state formalism entails  an 
evolution operator 
(see the following Eq (\ref{e3})) which is nonunitary and unbounded  
we,   nevertheless, show  that the results  obtained  are valid also for this 
kind of operator. That is, 
we obtain for the classical reactions  the same results that were 
 obtained \cite{Bar00}  for the 
quantum analogues  which are discussed  by using  complex coherent state
formalism  
\cite{Glauber63}, \cite{Klauder68} which  involves  unitary and bounded 
evolution operators. 
Needless to say that one may  use the conventional 
complex coherent state formalism \cite{Glauber63},  \cite{Klauder68} 
for discussing  classical 
reactions.  
\par
In Section 2 we discuss the reversible reaction 
$A_1+A_2+\cdots+ A_r\leftrightarrow B_1+B_2+\cdots +B_s$  for 
the special cases of $r=s=1$ and $r=s=2$.  That is, we study the
reversible reactions $A \leftrightarrow B$ and $A_1+A_2 \leftrightarrow 
B_1+B_2$, and show that repeating either direction of 
each   
 a large number $N$ of times  in a finite total time $T$ results, in the
limit of very large $N$,   in a unity probability to remain with  the initial 
reacting particles only.   The generalization to any finite $r$ 
and $s$ follows.   \par 
We  note that since we discuss  
  the probability to remain with  the initial reacting 
 particles the product particles of such reactions are not relevant (as the 
 reacting ones) to our
 discussion. In Section 3 we discuss the more general and natural case in which
 the product particles are relevant. That is, 
 we assume that the  particles of the  ensemble  interact  at
 different places and  times and that they begin from some given initial
 configuration of reactions and end at a final {\it different} one. 
 We assume that there are large number 
 of different
 paths of reactions that all begin at the given initial configuration and  end 
 at the  final one and  we calculate the probability that 
   a specified system of 
 reacting
 particles evolves along some  prescribed path of them. 
  We note that such paths of
``states'' for the diffusion controlled reactions have been discussed in
 \cite{Masao76}, \cite{Mikhailov81}, \cite{Mikhailov85},  \cite{Namiki92} 
 where  use was made of   quantum field 
 theory methods 
 \cite{Mahan93}, \cite{Enz92}, \cite{Mattuck76}, including 
 Wick's theorem \cite{Mahan93}, \cite{Enz92}, \cite{Mattuck76}, to derive 
 the classical 
 Feynman diagrams and paths \cite{Feynman48}, \cite{Feynman65}. 
   These methods were
 also used in \cite{Mikhailov81}, \cite{Mikhailov85} for chemical kinetics. We show that
 taking the limit of a very large number $N$ of  reactions  
 along  the prescribed path and performing them in a dense manner one obtains 
 a unity value for the 
 probability of evolution along  that path. \par 
 In Section 4 we use a numerical 
 model that has been used in \cite{Bar01} for showing the effect of dense
 measurement on classical systems.  This is 
 the model of two dimensional concentric billiard \cite{Bar01} 
  that is used here to  numerically 
 simulate  the reversible  reaction 
 $A+B \leftrightarrow A+C$.  
  The two possible modes of reflections inside the billiard;  either between 
  the two
 concentric circles or between points of the outer circle, represent the two
 directions of the reaction. We note that nuclear and radioactive reactions are
 well simulated by billiards in which the stationary scattering circles
 represent the interactions between particles (see, for example,  
 \cite{Bauer90}
 in which a model of a rectangular billiard with a circle inside was used to
 discuss the decay law of classical systems).    
 We show that the numerical billiard simulations  conform to the 
  analytical results   of 
 Sections 2 and 3.   
  \protect \section{The reversible reaction  
  $A_1+A_2+\cdots A_r \leftrightarrow B_1+B_2+\cdots +B_s$ \label{sec2}}
   We assume that 
we have some set of $N$ identical particles so that the configuration in
which the $i$th particle is located at $q_i$ $(i=1,2,\cdots,N)$ is defined as 
a
state of the system and denoted, in Dirac's notation, 
$|q_1,q_2,\cdots,q_N\!>$ 
    ($|q^N\!>$) (for distinguished sets of particles we partition \cite{Masao76}
    the total 
    $N$ system to 
    $N_1, \ N_2, \ \ldots$ subsystems). Thus,  when representing, in the 
    following, classical particles by
states we mean that they are elements of some configuration of the whole 
system.
Following this terminology we may calculate the probability to find the set 
of particles in some definite state $|q^N\!>$ as \cite{Masao76} 
$$F^{(N)}(q_1,q_2,\cdots,q_N;t)=\sum_{all  \ permutations \ of \ q_i}
f^{(N)}(q_1,q_2,\cdots,q_N;t), $$ where  $f^{(N)}(q_1,q_2,\cdots,q_N;t)$  is some
normalized distribution function. To this probability one assign, as done in 
  \cite{Masao76},  
a ``state''  $|F(t)\!>=\sum_{N=0}^{\infty}\int dQ^N
F^{(N)}(q_1,q_2,\cdots,q_N;t)|q^N\!>$, where $\int dQ^N=\frac{\int dq^N}{N!}$ (the
division by $N!$ is necessary \cite{Masao76} so as not to overcount the state $|q^N\!>$ $N!$ 
times). Thus,  the former probability to find the system in the state $|q^N\!>$ 
may be written as \cite{Masao76} $F^{(N)}(q_1,q_2,\cdots,q_N;t)=<\!q^N|F(t)\!>$.
\par  
We discuss first the general reversible reaction  
  $A_1+A_2+\cdots A_r \leftrightarrow B_1+B_2+\cdots +B_s$  for the 
  specific case of $r=s=1$,  
 $A \leftrightarrow B$  where $A$ and $B$ are, as noted, represented  
by the two coherent states \cite{Klauder68}
\begin{eqnarray}
&&|z_A\!>=e^{-\frac{1}{2}|z_A|^2}\sum_{n=0}^{n=\infty}\frac{z_A^n}{(n!)^{\frac{1}{2}}}|n\!>
\label{e1} \\
&&|z_B\!>=e^{-\frac{1}{2}|z_B|^2}\sum_{n=0}^{n=\infty}\frac{z_B^n}{(n!)^{\frac{1}{2}}}|n\!>
\nonumber  \end{eqnarray}
 Using the
following general equation for any two operators $X$ and $Y$  \cite{Klauder68} 
$$e^YXe^{-Y}=X+[Y,X]+\frac{1}{2!}[Y,[Y,X]]+\cdots, $$ where $[Y,X]$ is the 
commutation 
$[Y,X]=YX-XY$,  one obtains 
\begin{equation} \label{e2} U(q,p)(\alpha P+\beta
Q)U^{-1}(q,p)=\alpha(P+p)+\beta(Q+q)  \end{equation}
 The  $\alpha$,
$\beta$ are arbitrary parameters,  $U(q,p)$ and $U^{-1}(q,p)$ are given 
respectively by $U(q,p)=e^{pQ-qP}$, $U^{-1}(q,p)=U(-q,-p)$, and $Q$,
$P$ are the coherent state operators that satisfy $[Q_i,P_j]=\delta_{ij}$. 
That is, $U(q,p)$ translates the operators $Q$ and $P$ by $q$ and $p$
respectively.  Now, since the coherent states have been defined, as remarked, in
terms of the number representation eigenstates (see Eq (\ref{e1})) we 
 write the time evolution operator of the relevant states 
 as $e^{Nt}$,  where $N$ is the number operator
\cite{Klauder68} (note that since we discuss in this paper real coherent states 
 the evolution operator is real  also).  
$$N=a^{\dagger}a=(\frac{Q-P}{\sqrt 2})(\frac{Q+P}{\sqrt 2})=
\frac{1}{2}(Q^2-P^2+1)$$ 
$N$ is defined analogously to the corresponding operator of the complex coherent 
state formalism \cite{Glauber63},  \cite{Klauder68} but without the complex notation $i$ in the 
middle expression. Note that the operator $N$ is not positive definite and this 
to remind us, as remarked, that the real coherent state formalism discussed here 
does not conform to the conventional quantum operator process. Nevertheless, as 
remarked, the final results obtained here are exactly the same as those accepted 
in \cite{Bar00} from the quantum complex coherent state formalism. 
The commutation  
$[Q_i,P_j]=\delta_{ij}$ have been used at the right hand side of 
 the last equation. 
 Applying
the operator $N$ on the coherent state $|z\!>$ from Eq (\ref{e1}), 
and taking the scalar product of the
result with the conjugate  state  
$<\!z|$  one obtain (using
$<\!n|e^{Nt}|m\!>=e^{nt}\delta_{nm}$, since in the number representation the
operator $N$ is diagonal) 
\begin{eqnarray}
&& <\!\grave z|e^{Nt}|z\!>=\exp(-\frac{1}{2}|z|^2-\frac{1}{2}|\grave z|^2)
\sum_{n=0}^{n=\infty}\frac{(\grave
ze^tz)^n}{n!}=\nonumber \\ &&=\exp(-\frac{1}{2}|z|^2 -\frac{1}{2}|\grave z|^2+\grave
ze^tz) =<\!\grave z|e^tz\!>=\label{e3} \\ &&=<\!\grave z|(\cosh t+\sinh t)z\!>=<\!\grave q,\grave
p|q_t,p_t\!> \nonumber \end{eqnarray}   
  The last result is
obtained by writing $z$ in terms of $q, p$ in which we have  
\begin{equation}  q_t=q(\cosh t+\sinh t), \ \ \ p_t=p(\cosh t+\sinh t) 
\label{e4} \end{equation}  
  We, now,  calculate, using Eq (\ref{e3}),  the probability $p(|q_A,p_A\!>)$ to remain with the  initial particle $A$
 after the reaction $A \to B$ where the particle $B$ is represented by the 
 coherent state $e^{Nt}|z_A\!>$.  This is given by
\begin{eqnarray} 
&& <\!z_A|e^{Nt}|z_A\!>=<\!q_A,p_A|q_{A_t},p_{A_t}\!> 
=\exp(-\frac{1}{4}(q_A+p_A)^2- \nonumber \\ && -\frac{1}{4}(q_{A_t}+p_{A_t})^2)
\sum_{m,n=0}^{m,n=\infty}\frac{(q_A+p_A)^m(q_{A_t}+p_{A_t})^n}{2^{\frac{m+n}{2}}
(m!n!)^{\frac{1}{2}}}<\!m|n\!> =\nonumber \\ && 
=\exp(-\frac{1}{4}(p_A+q_A)^2- \frac{1}{4}(p_{A_t}+q_{A_t})^2)
\sum_{n=0}^{n=\infty}\frac{(q_A+p_A)^n(q_{A_t}+p_{A_t})^n}{2^nn!}
=\nonumber \\ && =\exp(-\frac{1}{4}(p_A+q_A)^2 -\frac{1}{4}(p_{A_t}+
q_{A_t})^2+ 
\frac{1}{2}(q_A+p_A)\cdot \label{e5} \\ && \cdot (q_{A_t}+p_{A_t}))
 =
\exp(-\frac{1}{4}(p_A+q_A)^2-\frac{1}{4}(p_A+q_A)^2(\cosh t+\sinh t)^2+
\nonumber \\ && +
\frac{1}{2}(q_A+p_A)^2\cdot (\cosh t+\sinh t))
 =\exp(-\frac{1}{2}(p_A+q_A)^2(\frac{1}{2}+\nonumber \\ && +\frac{1}{2}(\cosh t+\sinh t)^2 
 -(\cosh t+\sinh t))) \nonumber \end{eqnarray}
 Note that since we discuss coherent states the
interpretation \cite{Klauder68} of the expression $<\!z_A|e^{Nt}|z_A\!>$ is 
the probability to
find the {\it mean}  position and momentum of the coherent state $e^{Nt}|z_A\!>$,  
which represents $B$, equal to those of $z_A$, which represents $A$,  and this 
probability is equivalent 
in our discussion to remaining with the particle $A$. 
From Eq (\ref{e5}) one obtains  the probability to remain with the initial 
particle $A$ after a 
single reaction $A \to B$. If it is repeated $n$ times in a finite total
time $T$ one obtains (using $n=\frac{T}{\delta t}$, where $\delta t$ is the 
duration of each reaction)
\begin{eqnarray} && p^n(|q_A,p_A\!>)=\exp(-\frac{T}{2\delta t}
(p_A+q_A)^2(\frac{1}{2}+\frac{1}{2}(\cosh \delta t+\label{e6} \\ && +
\sinh \delta t)^2-
(\cosh \delta t+\sinh \delta t))) \nonumber \end{eqnarray}
In the limit of very large $n$ (very small $\delta t$) we expand the
hyperbolic functions in a Taylor series and keep terms up to second order in
$\delta t$.  We obtain 
\begin{eqnarray} &&
p^n(|q_A,p_A\!>)=\exp(-\frac{T}{2\delta t}(p_A+q_A)^2(\frac{1}{2}+
\frac{1}{2}(1+2\delta t^2+2\delta t)-\nonumber \\ &&-(1+\frac{\delta t^2}{2}+\delta t))
=\exp(-\frac{T}{4\delta t}(p_A+q_A)^2\delta t^2)= \label{e7} \\ && =
\exp(-\frac{T}{4}(p_A+q_A)^2\delta t \nonumber
\end{eqnarray}
Thus,  in the limit $n\to \infty$ ($\delta t\to 0$) we have 
\begin{equation} \label{e8} \lim_{n \to \infty} p^n(|q_A,p_A\!>)=\lim_{\delta t \to 0}
\exp(-\frac{T}{4}(p_A+q_A)^2\delta t)=1 \end{equation}
Now, although we refer in the former equations to  the direction $A \to B$ all 
our discussion remains valid also for the opposite one $B \to A$.  
That is, repeating either side of the  reaction $A \leftrightarrow B$ a large 
number of times $n$ in a finite
total time $T$ results, in the limit of very large $n$,  in remaining 
(with probability 1) with the initial 
particle of the repeated  reaction.\par
We, now, discuss the reversible 
reaction $A+B \leftrightarrow C+D$ in which we have two
 reacting particles.   We continue to use the number evolution operator $N$ 
  and take
into account that the initial particles $A$ and $B$ interact. Thus, representing
these particles as the coherent states $|q_{A},p_{A}\!>$ and 
$|q_{B},p_{B}\!>$ we write, for example,   the left hand side direction 
of the former reversible 
reaction $A+B \to C+D$ as 
 \begin{equation} \label{e9} \exp((N_A+N_B+P_AP_B+Q_AQ_B)t)|q_A,p_A\!>|q_B,p_B\!>=
 |q_C,p_C\!>|q_D,p_D\!>,  \end{equation}
 where the terms $Q_AQ_B$ and $P_AP_B$ represent,  as for the boson particles 
 discussed in    \cite{Klauder68},  
 the interaction of the
 particles $A$ and $B$, and $N_A$, $N_B$ are the number operators for them. 
 Note that, as for the reaction $A \leftrightarrow B$ (see the discussion after
 Eqs (\ref{e4}) and (\ref{e5})), the operation of the evolution operator on the
 coherent state $|q_A,p_A\!>|q_B,p_B\!>$, which
 is now more complicated due to the interaction between $A$ and $B$,  
 is represented by 
 $|q_C,p_C\!>|q_D,p_D\!>$. 
 We calculate, now, the probability that the reaction $A+B \to C+D$ results 
 in remaining with the initial particles $A$ and $B$ only  (we denote this
 probability by $p(|q_B,p_B\!>|q_A,p_A\!>)$). 
\begin{eqnarray} 
&& p(|q_B,p_B\!>|q_A,p_A\!>)= \label{e10} \\ && = 
<\!q_B,p_B|<\!q_A,p_A|\exp((N_A+N_B+P_AP_B+Q_AQ_B)t)|q_B,p_B\!>|q_A,p_A\!>
 \nonumber \end{eqnarray}
 Using Eqs (\ref{e1}), (\ref{e3})  and the following coherent states 
 properties \cite{Klauder68}
 $<\!q,p|Q|q,p\!>=q$, \ \ \  \  $<\!q,p|P|q,p\!>=p$ (derived  by using the operator $U$ 
  from Eq (\ref{e2})
 and the relation  $N|0,0\!>=0$)
we obtain 
\begin{eqnarray}
&&
p(|q_B,p_B\!>|q_A,p_A\!>)=\exp((q_Aq_B+p_Ap_B)t)\cdot \nonumber \\&& \cdot 
<\!q_B,p_B|<\!q_A,p_A|q_{B_t},p_{B_t}\!>
|q_{A_t},p_{A_t}\!>= 
\exp((q_Aq_B+p_Ap_B)t)\cdot \nonumber \\ && \cdot \exp(-\frac{1}{4}(q_A+p_A)^2-\frac{1}{4}(q_B+p_B)^2
-\frac{1}{4}(q_{A_t}+p_{A_t})^2- \frac{1}{4}(q_{B_t}+p_{B_t})^2)\cdot \nonumber
\\ && \cdot 
\sum_{m,n=0}^{m,n=\infty}\frac{(q_A+p_A)^m(q_{A_t}+p_{A_t})^n}
{2^{\frac{m+n}{2}}(m!n!)^{\frac{1}{2}}}<\!m|n\!>\sum_{s,r=0}^{s,r=\infty}
\frac{(q_A+p_A)^s(q_{A_t}+p_{A_t})^r}
{2^{\frac{s+r}{2}}(s!r!)^{\frac{1}{2}}}\cdot \nonumber \\ && \cdot 
<\!s|r\!>=
\exp((q_Aq_B+p_Ap_B)t)\exp(-\frac{1}{4}(q_A+p_A)^2-\frac{1}{4}(q_B+p_B)^2
-\nonumber \\ && -\frac{1}{4}(q_{A_t}+p_{A_t})^2-\frac{1}{4}(q_{B_t}+p_{B_t})^2)
\sum_{n=0}^{n=\infty}\frac{(q_A+p_A)^n(q_{A_t}+p_{A_t})^n}
{2^nn!}\cdot \label{e11} \\ &&  \cdot \sum_{r=0}^{r=\infty}
\frac{(q_A+p_A)^r(q_{A_t}+p_{A_t})^r}
{2^rr!} =\exp((q_Aq_B+p_Ap_B)t)\exp(-\frac{1}{4}(q_A+p_A)^2-\nonumber \\ && -
\frac{1}{4}(q_B+p_B)^2
-\frac{1}{4}(q_{A_t}+p_{A_t})^2 -\frac{1}{4}(q_{B_t}+p_{B_t})^2+ 
\frac{1}{2}(q_A+p_A)(q_{A_t}+p_{A_t})+\nonumber \\ &&+ 
\frac{1}{2}(q_B+p_B)(q_{B_t}+p_{B_t}))=
\exp(\frac{1}{2}(\cosh t+\sinh t)((q_A+p_A)^2+(q_B+p_B)^2)-\nonumber \\ && -
\frac{1}{4}(1+(\cosh t
+\sinh t)^2)\cdot  ((q_A+p_A)^2+  (q_B+p_A)^2)+ (q_Aq_B+p_Ap_B)t) 
\nonumber \end{eqnarray}
This is the probability to remain with the original particles $A$ and $B$ after
one reaction. Repeating it a large number of times $n$ in a finite total time $T$,
where $n=\frac{T}{\delta t}$ ($\delta t$ is the time duration of one reaction) 
one  obtains for the probability to remain with  $A$ and $B$. 
\begin{eqnarray} && P^n(|q_B,p_B\!>|q_A,p_A\!>)=\exp(\frac{T}{\delta t}
((q_Aq_B+p_Ap_B)\delta t+\nonumber \\ && + 
(\frac{1}{2}(\cosh \delta t+\sinh \delta t)
-\frac{1}{4}(1+(\cosh \delta t+\sinh \delta 
t)^2))\cdot \label{e12} \\ && \cdot ((q_A+p_A)^2+(q_B+p_B)^2))) 
\nonumber \end{eqnarray}
In the limit of very large $n$ we expand the hyperbolic functions in a Taylor
series and retain terms up to second order in $\delta t$. Thus,  
\begin{eqnarray} && P^n(|q_B,p_B\!>|q_A,p_A\!>)=\exp(\frac{T}{\delta t}
((q_Aq_B+p_Ap_B)\delta t+
(\frac{1}{2}(1+\frac{\delta t^2}{2}+\nonumber \\ && + \delta t)
-\frac{1}{4}(2+2\delta t^2+2\delta t))((q_A +p_A)^2
 +(q_B+p_B)^2)))=\label{e13} \\ && =
\exp(T((q_Aq_B+p_Ap_B)- 
\frac{\delta t}{4}((q_A+p_A)^2+(q_B+p_B)^2))) \nonumber 
\end{eqnarray}
Taking the limit of $n\to \infty$ ($\delta t \to 0$) we obtain 
\begin{equation} \label{e14}  \lim_{n\to \infty}P^n(|q_B,p_B\!>|q_A,p_A\!>)= 
\exp(T(q_Aq_B+p_Ap_B)) \end{equation}
The last  probability tends to unity when the $c$-numbers of either
$A$ or $B$ (or both) are zeroes, that is, when $A$ or $B$ (or $A$ and $B$) 
  are in their ground states (in which case they are    
  represented by only the first term of the sums in Eqs
 (\ref{e1})).  
Needless to say  that all the former
 discussion  remains valid also for  the opposite direction 
 $A+C \to A+B$.  Thus,  we conclude that when either direction of the 
 reversible reaction $A+B
 \leftrightarrow C+D$ is   repeated
a large number  of times $n$  in a finite total  time and when at least 
one of the 
reacting particles 
was in the  ground state so that its  $c$-numbers 
 are zeroes  one obtains, in the limit of very large $n$,   a result as if 
 the repeated reaction  did  not occur at all.    \par 
 It can be
shown that the general reversible reaction 
 $A_1+A_2+\cdots A_r \leftrightarrow B_1+B_2+\cdots +B_s$, 
where $r$, $s$ are any two 
positive natural numbers,  also
results in a similar outcome if at least one of
the reacting particles has zero $c$ numbers. We note that  the last  
condition  is not necessary when we begin with only one reacting particle
as we see from the reaction $A \leftrightarrow B$. 
\protect \section{ The probability to find given final configuration  
 different from the initial one \label{sec3}} \noindent 
 We now discuss the more general and natural case in which we 
have an ensemble of particles and we 
calculate the
probability to find at the time $t$ a subsystem of this ensemble 
at some given
configuration if at the initial time $t_0$ it was at another prescribed 
 configuration.  We 
assume that the corresponding time difference  
$(t-t_0)$ is not
infinitesimal and that during this time  the subsystem has undergones a series of 
reactions.  The passage from some reaction at some intermediate 
time $t_i$ to the neighbouring one at the time $(t_i+\delta t)$ 
is governed by the correlation  between the corresponding resulting configurations 
of the subsystem at these times. 
Thus, restricting, for the moment, our
attention to the case in which a particle  in the subsystem that was  
 at the time $t_0$ in  the state $A$ and  at the time $(t_0+\delta t)$ 
 in    $B$   we can write    the relevant correlation \cite{Klauder68}  
 between these two states as 
\begin{equation} \label{e15} \tau(A,B;t_{0_A},(t_0+\delta t)_B)=
<\!q_{A_{t_0}},p_{A_{t_0}}|q_{B_{t_0+\delta t}},p_{B_{t_0+\delta t}}\!>,
  \end{equation}  where  $|q_{A_{t_0}},p_{A_{t_0}}\!>$  and 
  $|q_{B_{t_0+\delta t}},p_{B_{t_0+\delta t}}\!>$
  are the coherent states that represent the particles A 
and B 
at the times $t_0$ and  $t_0+\delta t$ respectively (see Eqs (\ref{e3}),(\ref{e4})) 
 and the angular brackets denote 
an   ensemble average over all the particles of it. We note
that if $A=B$, $\tau$ measures \cite{Klauder68} the autocorrelation of
either the particle A or B, and when $A \ne B$  $\tau$ is the
crosscorelation \cite{Klauder68} of the two particles. It can be shown,
using Eqs (\ref{e1}) and (\ref{e15}) that the following relation  
\begin{equation} |\tau(A,B;t_{0_A},(t_0+\delta t)_B)|^2=\tau(A,A;t_{0_A},t_{0_A})
\tau(B,B;(t_0+\delta t)_B,(t_0+\delta t)_B),  \label{e16} \end{equation} 
is valid. That is, the modulus of the crosscorrelation of the particles A and B 
 at the times $t_0$ and $(t_0+\delta t)$  equals the product of the
 autocorrelation of the particle $A$ at the time $t_0$ by that  
 of  $B$ at the time $(t_0+\delta t)$. The last relation is interpreted
 \cite{Klauder68} as the probability density for the occurence of 
 the reaction  $A \to B$ at the time
 $(t_0+\delta t)$. That is, given that the system was in 
 ``state'' $A$ at the time $t_0$, the probability to find it at the time 
 $(t_0+\delta t)$ 
 in  ``state'' $B$ is given by Eq (\ref{e16}). We can generalize to the joint
 probability density for the occurence of $n$ different reactions  between the 
 initial  and  final times $t_0$ and   $t$, where each   involves  
  two different particles and is of the kind $A_i \to  A_{i+1}$. That is, each
  reaction is composed of two parts; the first one is that in which a particle of the
  subsystem  is observed at the time $t_i$ to be in state $A_i$, and the second
  that in
  which it is observed at the time $t_i+\delta t$ to be  in  the state $A_{i+1}$.
  Thus, the total time interval $(t-t_0)$ is partitioned into $2n$ subintervals
  during  which the $n$ reactions occur.    The total correlation  is  
  \begin{eqnarray} && |\tau(A_1,A_2,\cdots,A_{2n};t_0,t_0+\delta t,\cdots,t)|^2=
 \tau(A_1,A_1;t_0,t_0)\cdot \nonumber \\ && \cdot 
 \tau(A_2,A_2;t_0+\delta t,t_0+\delta t)\cdots 
   \tau(A_{2n},A_{2n};t,t)= \label{e17} \\&& =
 \prod_{k=0}^{k=2n-1}\tau(A_{k+1},A_{k+1};t_0+k\delta t,t_0+k\delta t)
 = \prod_{k=0}^{k=n-1}\tau(A_{2k+1},A_{2k+1};t_0+\nonumber \\ && +
 2k\delta t,t_0+2k\delta t)
 \cdot \tau(A_{2k+2},A_{2k+2};t_0+(2k+1)\delta t,t_0+(2k+1)\delta t)= \nonumber \\ 
&& =\prod_{k=0}^{k=n-1}
|\tau(A_{2k+1},A_{2k+2};t_0+2k\delta t,t_0+(2k+1)\delta t)|^2 \nonumber 
 \end{eqnarray}
The last result was obtained by using 
Eq (\ref{e16}).  By the
notation $A_{2n}$ we mean, as remarked, 
that there are $n$ separate reactions  each of which involves two states 
(and not $2n$ different particles).   Now, 
we show in  the previous section,  for either direction of the
reversible reaction
$A \leftrightarrow B$,  that the probability  
to remain in the initial  state $A$  (or $B$) 
tends to unity in the limit of a very large number of repetitions, 
in a finite total time, 
of $A \to B$ (or $B \to A$) 
which amounts to performing each such reaction in an infinitesimal
time $\delta t$.  That is, in this limit of vanishing $\delta t$  
each factor of the last product in 
Eq (\ref{e17}),  
which is  the probability for the reaction  $A_i \to  A_{i+1}$, 
tends 
to unity and with it the joint probability for the occurence of the $n$
reactions. Thus,     
     the specific
prescribed path of reactions  is followed   with a probability
of unity. 
\par 
From the last discussion we  may obtain the joint probability density for the
case in which some of the $n$ intermediate reactions are of the more general
kind $A_1+A_2+\cdots + A_r \to B_1+B_2+ \cdots +B_r$, where $r$ is an arbitrary
natural positive number. That is, at some of the $2n$  times  there may occur, 
in a simultaneous manner, $r$ different reactions at
$r$ different places each of the kind $A \to B$. Thus, we assume 
 that
$r$ particles in the subsystem  that were  at the time $t_0+i\delta t$ in  the
states $A_j$ ($j=1, 3, 5,\cdots,2r-1$) were  observed at 
the time $(t_0+(i+1)\delta t)$ to
be in  the states $A_{j+1}$ ($j+1=2, 4, 6,\cdots,2r$).  
 We assume that at each of 
the other
intermediate times there occurs only one 
 single reaction $A_i \to A_{i+1}$. 
Thus, there
are $(n+r-1)$ reactions each of them occurs  between two particles. In this
case the corresponding total coherence among all these reactions is \begin{eqnarray*}
&&\tau_{total}=\tau(A_1,A_2,\cdots,A_{i+1},A_{i+2},\cdots,A_{i+2r},\cdots,A_{(2n+2r-2)};t_0,t_0+\delta
t,\cdots \nonumber \\&& \cdots,\underbrace{t_0+i\delta t,t_0+(i+1)\delta t,\cdots,
t_0+i\delta t,t_0+(i+1)\delta t}_r,\cdots,t),  \nonumber \end{eqnarray*} 
where the underbrace denotes that the $r$ particles observed at the time
$(t_0+i\delta t)$ as 
$A_j$  were seen to be at the time $t_0+(i+1)\delta t$ as 
$A_{j+1}$ ($j=1, 3,\cdots,2r-1$).  
Again the notation $A_{(2n+2r-2)}$ means that we have
$(n+r-1)$ reactions each involving, as remarked, two states. 
Using Eqs (\ref{e16})-(\ref{e17}) and the former equation for $\tau_{total}$ we 
find that the joint probability 
to find   at the time $t$  
the relevant subsystem at the given final configuration after the occurence of 
 these  $(n+r-1)$
reactions 
   is   
  \begin{eqnarray} && |\tau_{total}|^2=\tau(A_1,A_1;t_0,t_0)\tau(A_2,A_2;t_0+  
\delta t,t_0+\delta
t) \cdots 
\tau(A_{i+1},A_{i+1};t_0+\nonumber \\ && +i\delta t,t_0+i\delta t) 
 \cdot \tau(A_{i+2},A_{i+2};
t_0+(i+1)\delta t,t_0+(i+1)\delta t)\cdots  \nonumber \\ && \cdots 
 \tau(A_{i+2r-1},A_{i+2r-1};t_0+i\delta t,t_0+i\delta t)  
 \cdot \tau(A_{i+2r},A_{i+2r};
t_0+(i+1)\delta t,t_0+\nonumber \\ && + (i+1)\delta t) \cdots
 \tau(A_{(2n+2r-2)},A_{(2n+2r-2)},t_0+
 2n\delta t,t_0+2n\delta t)= \nonumber \\ && =  
\prod_{k=0}^{k=n-1}\tau(A_{2k+1},A_{2k+1};t_0+2k\delta t,t_0+2k\delta t)
\tau(A_{2k+2},A_{2k+2};t_0+\nonumber \\ && +(2k+1)\delta t,t_0+
(2k+1)\delta t) \cdot 
\prod_{j=1}^{r-1}\tau(A_{i+j},A_{i+j};t_0+i\delta t,t_0+i\delta t) 
\cdot  \label{e18} \\ && \cdot \tau(A_{i+j+1},A_{i+j+1};t_0+(i+1)\delta t,t_0+
(i+1)\delta t)= 
\prod_{k=0}^{k=n-1}
|\tau(A_{2k+1},A_{2k+2};t_0+\nonumber \\ && + 2k\delta t,t_0+(2k+1)\delta t)|^2  
 \prod_{j=1}^{j=r-1}|\tau(A_{i+j},A_{i+j+1};t_0 +i\delta t,t_0+(i+1)\delta t)|^2 
  \nonumber  \end{eqnarray}
   The first product of the last result is the same as that of Eq (\ref{e17}) 
   and the second takes
 account of  the $r-1$ simultaneous reactions at the time $(t_0+(i+1)\delta t)$ 
 (the first product involves also one of the $r$ simultaneous reactions at the  
 time $(t_0+(i+1)\delta t)$). Each of the reactions in both products is of the kind
 $A \to B$ which was  shown in the former section (see also the discussion
 after Eq (\ref{e17})) to have a unity  probability for  remaining 
  with the initial particle
 $A$  in the limit in which the time duration of the reaction  becomes
 infinitesimal. That
 is, in this limit in which the time alloted for each reaction $A \to B$ 
 becomes very small  each  factor of each product of Eq (\ref{e18}), and
 with it the whole expression, tends to unity. If any particle $A$ of the 
 subsystem 
 does not react with any other particle at some of the $2n$ intermediate times
 then we may denote its no-reaction at these times as $A \to A$ and the
 probability for it to remain in the initial state (which is the same as the final
 one) is obviously unity.
   \par 
  Thus, we see that the probability to find at the time $t$ the given ensemble 
of particles  following 
a given path of reactions (from a large number of possible paths) 
tends to unity if the relevant  reactions 
are performed in a dense manner.  That is, each occuring in an infinitesimal
time $\delta t$. 
 \protect \section{\bf  Billiard simulation of the reversible reaction $A+B
 \leftrightarrow A+C$ \label{sec4}} 
 \noindent
 We, now,  show 
  that the  results   described  
 in the former sections  have also strong numerical support.    
  This is demonstrated  for  the reversible reaction $A+B 
 \leftrightarrow A+C$  
 which is  simulated  here   by using 
  the two-dimensional circular billiard \cite{Bar01} which is
 composed of two concentric circles. We assume that  
  initialy we have a large ensemble of
 identical point particles each of them is the component $A$ of the  
 reaction  $A+B 
 \leftrightarrow A+C$. All of   these particles are  entered, one at a time, 
   into the billiard  in which they are elastically reflected by the two
   concentric circles. That is, the angles before and after each reflection are
   equal. We assume that on the outer circle there is a narrow hole through
   which the particles $A$ leave the billiard. Once a particle is ejected
   out a new one is entered and reflected inside the billiard
   untill it leaves and so  for all the particles of the ensemble. There are
   two different possible kinds of motion for each point particle $A$ before 
   leaving the
   billiard; either it is
   reflected between the two cocentric circles or, when the angle of reflection
   is large, reflected by the outer circle only without touching the inner one.
   Now, since both motions  are elastic each particle $A$,  once it 
   begins its reflections  in
   either kind of motion,    continues to move   only in this kind     
    untill it leaves through the narrow opening. The
   component $B$ of the reaction denotes the outer larger circle, and the
   component $C$ denotes both circles. That is, the left  hand side  $A+B$ of the 
   reaction  
   signifies that the point particle $A$ moves inside the billiard and is 
   reflected by the outer circle only, whereas,  the right  hand side   
   $A+C$ denotes the second kind of motion in which the point balls $A$ are reflected
   between the two circles.  We call these two kinds of paths ``states''
   \cite{Bar01}, so that the path that
   touches both circles is ``state'' 1 and the one that thouches the outer circle only
   is ``state'' 2. This billiard model was studied in \cite{Bar01} as an
   example of a classical system that behaves  the same way   
    quantum systems  do  when exposed to a large number of repetitions, in a
   finite total time, of the same experiment 
   (\cite{Misra77}, \cite{Giulini96}, \cite{Pascazio94}, \cite{Cook88}, 
   \cite{Harris81}, \cite{Bixon82}, \cite{Itano90}, \cite{Kofman96}, 
   \cite{Simonius78}, \cite{Aharonov80},  \cite{Facchi99}). The concentric billiard is,
   schematically, represented in Figure 1. 
   
   \begin{figure}[hb]
\centerline{
\epsfxsize=3in
\epsffile{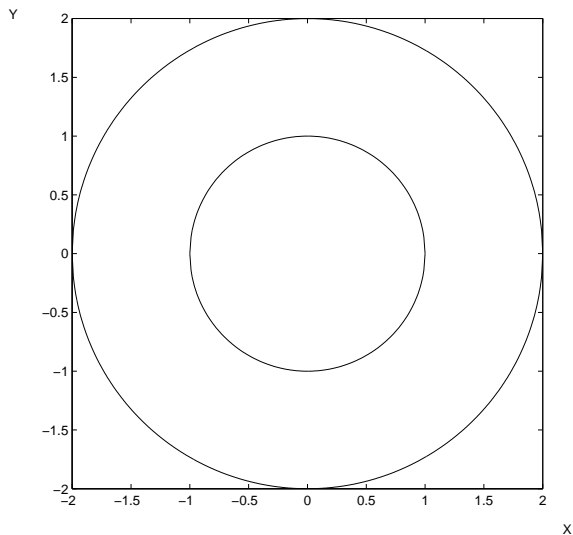}}
\caption[fegf114]{A schematic representation of the concentric circular 
billiard that simulates the reversible reaction $A+B \leftrightarrow C+D$.}
\end{figure}

   \par  Now, since in such a system we 
   can not
   follow the path of each particle and can not differentiate  between the two kinds 
   of motion  we have to consider, as done for the nuclear and radioactive 
   processes \cite{Bauer90},
    the activities of these particles in either path.  
     That is, the rate at which
    the entire ensemble of particles, being at either  state,   
     leaves the
    billiard. We assume for the activity discussed here, as is assumed
    \cite{Bauer90} for the nuclear and radioactive's activities, that each particle $A$
    enjoys arbitrary initial conditions, so in the following numerical
    simulations we assume that it may begin its journey inside the billiard at
    either ``state'' which is determined  randomly using a random number 
    generator. \par
     We want to show, 
    numerically,  that if either side of this reversible  reaction $A+B
    \leftrightarrow A+C$ is repeated a large number
     of times $N$ in a finite total time $T$, then,  in the limit of very large
    $N$,  the activity obtained is the same as the ``natural activity'' 
    \cite{Bar01} that  results
    when no such  repetitions are performed \cite{Bar01}. 
       For that matter, we 
    take into account that the reversible reactions that  occur 
    in nature have either equal or different rates for  the two
    directions of the reactions and that the total activity of the ensemble 
     of particles  depends critically upon these
    rates \cite{Bauer90}. 
     If, for example, we consider the equal rate case 
     then we have to discuss   the rate of evacuation
    of the billiard when each particle  
    is allowed,  after a  prefixed number of reflections in either state,  
     to pass, if it is still in the billiard,
    to  the other one.  This  activity is shown by the solid curve in 
      Figure 2 in which the
    ordinate axis denotes the number of particles $A$ that leave the billiard in
    prescribed time intervals binned in units of 60 \cite{Bauer90}. We assume 
    \cite{Bauer90} that
    each point particle $A$ in either state moves with the same speed of 3, and
    the hole through which they leave has a width of 0.15. We denote the outer
    and inner radii of the billiard by $r_1$ and $r_2$ respectively, and 
    assign them the values of $r_1=6$ and $r_2=3$. The initial number of the
    particles $A$ was $10^6$, and each one of them passes from one ``state'' to the
    other, if it did not leave the billiard through the hole, after every 1100 
    consecutive reflections. We note that this rate of one passage for every
    1100 reflections is typical and common for these kinds of billiard
    simulations \cite{Bauer90}, \cite{Gutkin86}, \cite{Hobson75}. \par 
    The  natural activity is obtained,  as remarked,   when the entire ensemble 
    of $10^6$ particles
    $A$ enter, one at a time, to the billiard at the same  ``state'' and
    remain all the time in this ``state'' without passing to the other  until
    they leave the  billiard.  The dashed curve in Figure 2 shows this natural 
    activity  when  all the particles $A$ are in ``state''  2 in which they are 
    reflected only between points of the outer circle untill they leave the
    billiard. The dash-dot curve shows  the activity when all the particles $A$ 
    are in ``state'' 1  in
    which  they  are reflected only between the two circles. It has been found
    that for the values assigned here to the radii of the outer and inner
    circles (6 and 3) the activity of ``state'' 2 shown by the dashed curve is 
    the maximum
    avilable and that of state 1 shown by the dashdot curve is the minimum. The
    large difference between the two activities has its source in the range of
    the allowed angles of reflections which is much larger in state 2 than in 
    state 1. This is because the minimum trajectory between two neighbouring
    reflections in state 2, where the particles $A$ are reflected between points
    of the outer circle only, may be infinitesimal  compared to the
    corresponding trajectory in state 1 which is (we denote the trajectories 
    between
    neighbouring reflections in states 1 and 2 by $d_1$ and $d_2$ respectively)  
    $d_{1_{min}}=r_1-r_2$.    For
    the values assigned here to the radii $r_1$ and $r_2$ of the two concentric
    circles ($r_1=6$ and $r_2=3$) $d_{1_{min}}=3$. We note that the 
    maximum trajectory between two
    neighbouring reflections in state 2 is equal to the corresponding one in
    state 1, that is $$d_{1_{max}}=d_{2_{max}}=\sqrt{r_1^2+r_2^2}$$ 
    Thus, the particles in $A$ have  many  more possibilities to be reflected 
    to the
    hole and leave the billiard in state 2 than in state 1 
    and,  accordingly,   their activity is much 
    larger. The solid curve in Figure 2 is, as remarked, the activity
    obtained when the particles $A$ are transferred between the two states at
    the rate of one passage for every 1100 reflections and
    so, as expected, its activity  is between the two other activities 
    shown in Figure 2. 
    
\begin{figure}[hb]
\centerline{
\epsfxsize=3in
\epsffile{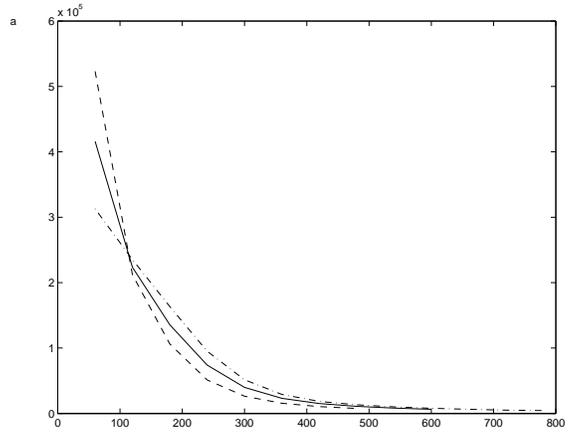}}
\caption[fegf004]{The   dashed   curve shows the activity obtained when all
the $10^6$  particles of the ensemble are allowed to be only in ``state'' 2,  
in which  they
are reflected  between points on the outer circle only. The dashdot curve is the
activity    when all these  particles  are allowed to move only in
``state''  1 (between the two circles). For the values
assigned here to the outer and inner circles (6 and 3) the dashed (dashdot) curve
is the maximum (minimum) activity. 
The solid curve shows the activity obtained when the particles in either state
pass to the other after every 1100 reflections. The $x$ axis specifies time
binned in units of 60. }
\end{figure}

    \par 
     We numerically  interfere  with the rate of the systematic passage of the 
     point
    particles $A$ between the two states such that this rate is accelerated. 
    We refer in the following not only to a specific passage between the two
    "states"
     but
    also gives the results in parentheses for the opposite passage. Thus, 
    it is found that the activity of the entire ensemble is directly (inversely)
    proportional to    the rate of the
    passage from state 1 (2) to state 2 (1) when the
    opposite passage from state 2 (1) to state 1 (2) remains at the  rate 
     of one  for every 1100
    reflections.  Thus, we have 
    found that when the particles in  state 1 (2) are transferred to  
    state 2 (1) 
     at the maximum rate of one passage for  each single reflection and the 
     particles in
      state 2 (1) are passed to the state 1 (2)   at the  
      rate  of one for every 1100
    reflections then the activity of the
     particles $A$ is maximal (minimal).  But as we have  remarked the
     maximal (minimal) activity is obtained only when each particle of 
     the entire ensemble is {\it always} in state 2 (1).  
     In other words, as we have  remarked,  a very large number of 
     repetitions of the left (right)  direction $A+B \to A+C$ ($A+C \to A+B$)  
     of the   reaction    where the   right (left) direction $A+C \to A+B$ 
     ($A+B \to A+C$)   occurs   every 1100 reflections,  yields a   result as  
     if the densely repeated  reaction never happened and the 
     activity obtained is  the natural  one in which no external 
     repetition is present. 
        The dashed curve in Figure 3, which is the same as the dashed one of
      Figure 2,  shows the activity obtained when all the $10^6$ particles $A$
      of the ensemble are allowed to move only in state 2 until they leave the
      billiard. The solid curve is the activity obtained when the reaction 
      $A+C \to A+B$ is repeated after each single reflection and the opposite
      one   $A+B \to A+C$ after every 1100 reflections. 
       It is seen that these  curves 
      are  similar to each other.  That is,  we realize, in accordance
     with the analytical results of the former sections,    that 
        a large number of repetitions of the 
     reaction
     yields a result that characterizes the activity obtained in the 
     absence of
     such repetitions.

     \begin{figure}[hb]
\centerline{
\epsfxsize=3in
\epsffile{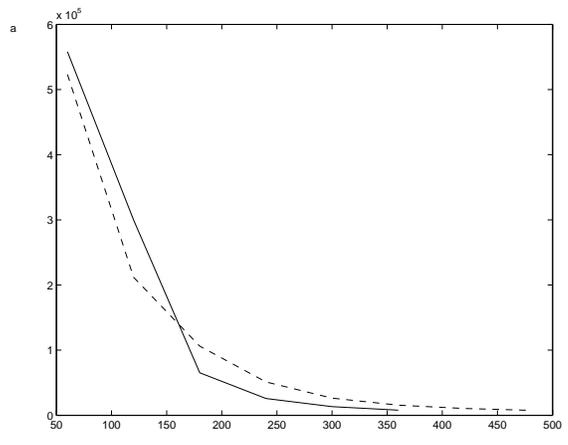}}
\caption[fegf14]{The  dashed  curve, which is the same as the dashed curve of
Figure 2 (they look slightly different since the abcissa axes of these figures
are different),  shows the activity obtained when all the
$10^6$ particles
$A$ of the ensemble are numerically constrained to be only  in ``state'' 2 
until they evacuate the billiard. State 1 is not allowed for them. The solid
curve is the  activity obtained when each particle in state 1 is passed 
 to  ``state'' 2  after each single reflection, whereas 
those in ``state'' 2  pass to the opposite one only after every 1100
reflections.  As for figure 2 the abcissa axis denotes time binned in units of
60.  Note the similarity between the two curves.   }
\end{figure}
     This is seen, in a much more clear way, in Figure 4 for 
     the
     other direction $A+B \to A+C$  of the reaction. The apparent  single graph
     of the figure is actually composed of two curves; one solid and the other
     dashed. The solid curve  shows the activity obtained
     when the reaction $A+B \to A+C$ is repeated after each single 
     reflection and  the
     opposite one $A+C\to A+B$ after 
     every 1100 reflections. The dashed curve,  which is identical to the 
     dash-dot
     one from Figure 2,  is the activity  obtained when all the particles $A$ of
     the ensemble are constrained to move only in state 1 until they leave the
     billiard. Note that the two curves are almost the same except for the
     longer tail of the dashed curve. 
     
     \begin{figure}[hb]
\centerline{
\epsfxsize=3in
\epsffile{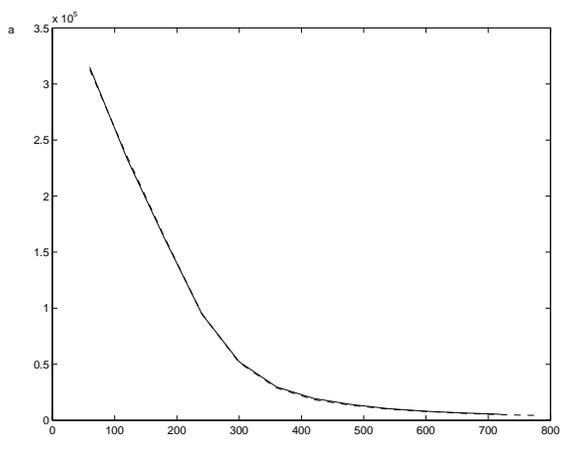}}
\caption[fegf04]{The apparently one curve shown in the figure is actually two
curves one dashed and the other solid. The  dashed  curve, which is the same 
as the dashdot curve of
Figure 2,  shows the activity obtained when all the
$10^6$ particles
$A$ of the ensemble are numerically constrained to be only  in ``state'' 1 
until they evacuate the billiard. State 2 is not allowed for them. The solid
curve is the  activity obtained when each particle in state 2 is passed 
 to  ``state'' 1  after each single reflection, whereas 
those in ``state'' 1  pass to the opposite one only after every 1100
reflections.  As for figure 2 the abcissa axis denotes time binned in units of
60.  Note that the two curves are almost identical (the dashed curve has a
longer tail (for large $t$) than the solid one).}
\end{figure}
     
     From both Figures 3 and 4 we realize that 
       the large number of repetitions of either direction  of the reversible 
      reaction $A+B \leftrightarrow A+C$ has the effect as if it has not 
      been performed at all and
     the actual activity obtained is  that of the natural one that does not
     involve any repetitions. \par
      We note that as the analytical results are obtained in the limit 
      of the  {\it largest}  number (actually infinite) of repetitions so the 
      similar 
      numerical results are obtained in 
       the limit of the largest  number of repetitions of
      the reaction.  That is, of numerically repeating it after each single
      reflection. In other words, a  mere high rate (which is not the maximal)
       of one side of the reaction   is not enough to
      produce the results shown in Figures 3-4. This is clearly shown in Figure
      5  the solid curve of which   shows the activity obtained
      when each particle in ``state'' 1 is passed to ``state'' 2 after 
      every {\it two} 
      neighbouring  reflections whereas those of ``state'' 2 are passed 
      (one at a
      time) after every 1100 reflections. Note that the solid curve in Figure 3
      shows the activity obtained when the particles in ``state'' 1 are passed
      to ``state'' 2 after each reflection and those of 2 passed to 1 after
      every 1100 reflections. That is, although the two high rates represented
      by  the two solid curves in 
      Figures 3 and 5 are almost the same nevertheless the resulting activities,
      contrary to what one may expect, are very different. That is, that of
      Figure 3 is much higher than that of Figure 5 as may be seen from the
      solid curve   that begins at $t=60$ (note that our  abcissa axis
      is binned in units of 60) from the high 
      value of $5.65 \cdot 10^5$ and ends at $t=360$. The corresponding solid
      curve of Figure 5 begins at $t=60$ at the much smaller value of $4.45
      \cdot 10^5$ and ends at the later time of $t=420$. That is, by only
      increasing the rate of repeating the same reaction from one for every two 
      reflections to one for each reflection results in an additional 120000
      particles that leave the billiard already at the first binned time unit. 
        The two dashed curves
       of Figures 3 and 5 are identical  and denote the same activity obtained
        when all the $10^6$ particles
$A$ of the ensemble are numerically constrained to be only  in ``state'' 2 
until they evacuate the billiard.

\begin{figure}[hb]
\centerline{
\epsfxsize=3in
\epsffile{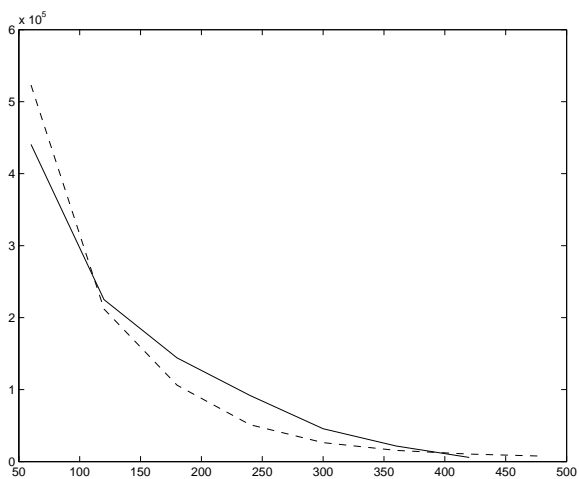}}
\caption[fegf05]{The  dashed  curve, which is the same as the dashed curve of
Figure 3,    shows the activity obtained when all the
$10^6$ particles
$A$ of the ensemble are moving inside the billiard  only  in ``state'' 2 
until they are evacuated outside. State 1 is not allowed for them. 
The solid
curve is the  activity obtained when each particle in state 1 is passed 
 to  ``state'' 2  after every  two reflection, whereas 
those in ``state'' 2  pass to the opposite one only after every 1100
reflections.  Note that although the two solid curves of Figures 3 and 5 
differ slightly in their  reaction  rates 
(of one for each reflection in Figure
3 and of one for each two reflections in Figure 5) 
 the activities are very different (see
text). 
As for all the former figures  the abcissa axis denotes time binned in units of
60.}
\end{figure}

 Thus, as remarked,  the important factor
that causes a result of maximum activity is the highest possible 
rate and not merely a large ratio between the higher and slower frequencies.       
      This is in accord with the analytical results  of  Sections 2 and 3
      which are obtained for   the largest rate (actually infinite) 
       of repeating any  direction of the
      general reversible reaction $A_1+A_2+\cdots A_r \leftrightarrow 
      B_1+B_2+\cdots +B_s$, 
      where $r$ and  $s$ are any two arbitrary natural positive numbers.   
      In this limit we find, as found from the billiard simulation, 
      that one  remains,   with a unity probability,  with 
      the initial reacting particles as if the  repeated reaction did not occur
      at all. \par  
      All the former simulations were done when the outer and inner 
     circles radii were 6 and 3 respectively. We note that we obtain similar
     numerical results for all other assigned values of  $r_1$ and $r_2$ up to 
     the extreme limits of $r_1>\!>r_2$ and $r_1 \approx r_2$ provided we always
     have $r_1>r_2$.  \par
     We may explain these results along the same line of explanation used to
   interpret the similar results obtained in the quantum regime. It have   been 
   established theoretically (\cite{Misra77}, \cite{Simonius78}, \cite{Aharonov80},
   \cite{Pascazio94}, \cite{Facchi99}) and experimentally (\cite{Itano90}, 
   \cite{Kofman96}, \cite{Giulini96}) that
   taking some quantum system which may reduce  by 
     experimenting on it  
    to any of its relevant 
   eigenstates  
     and repeat this  
    experiment a large number $N$ of times in a finite total time $T$ 
   results,  in the limit of very large $N$  (keeping $T$ constant), 
   in preserving the initial   state of 
   the system.
   This phenomenon is the Zeno effect (\cite{Misra77}, \cite{Simonius78}, 
   \cite{Aharonov80},
   \cite{Pascazio94}, \cite{Facchi99}). The similar results obtained
   theoretically in Section 2 suggest that this
    effect may be effective also in the classical reactions. That is, 
    repeating these reactions 
     a   large number of times, in a finite total time,  may results  in 
    remaining with
    the initial reacting particles  (initial "state").
    Moreover, it have been shown (\cite{Aharonov80}, \cite{Facchi99}) that if 
    any quantum
    system  evolves in a number of different possible paths of states,  each of
    which 
     begins  at the same given initial state and end at another given common
    final one,  then it is possible to realize any such path by making dense 
    measurement (in a finite total time) along it. 
    That is, by performing in a dense manner the set 
     of  experiments 
    that reduce the system to the specific states that constitute the relevant 
    path.  Similar results were obtained
    in Section 3, in which we show that the joint probability density for the 
    occurence of $n$ given  different  reactions between the initial 
    and final
    times $t_0$ and $t$  tends to unity in the limit of densely performing  
    these  reactions (when $n \to \infty$). \par  The same results  
    were  obtained    also in the billiard 
    simulations from which we realize that repeating a large number of times 
    any
    direction of the reversible reaction $A+B \leftrightarrow A+C$  have, in the
    limit of numerically repeating it   after each single reflection,   
    the effect as if it has never occured.   This  is what one obtains 
       in the Zeno effect  (\cite{Misra77}, \cite{Simonius78}, \cite{Aharonov80},
   \cite{Pascazio94}, \cite{Facchi99}, \cite{Itano90}) where  
    the
    system is preserved in the initial state in spite   of the very
    large number of measurements.  
 \protect \section*{\bf Concluding Remarks \label{sec5}} \noindent 
 We have discussed  classical reactions  using  quantum theory methods in 
 which particles are
represented by coherent state functions and the product of these states  
with their 
conjugates is
interpreted as probability \cite{Masao76}.   It is shown that if either 
direction of the general reaction  
$A_1+A_2+\cdots A_r \leftrightarrow B_1+B_2+\cdots +B_s$, 
where $r$ and  $s$ are any two natural positive numbers, 
is repeated a large number $N$ of times  in a finite total time then   in the 
limit of a very large $N$ one    remains,  with a unity probability,  
 with the
initial particles  only. In this context we
differentiate  
between the case in which
there were more than one initial reacting particle and the case in which 
there is  only one
such  particle. In the first case the mentioned unity probability is obtained 
 if at
least one of the initial reacting particles have zero values for 
its $c$-numbers
$q$ and $p$ that denote its coherent state representation, whereas, in the
second case this condition is not needed.   
 Moreover, it has been shown
that any  prescribed evolution (from a large number of possible ones) 
through a sequence of specific reactions  may be realized with
a unity probability by densely performing these reactions. 
This effect that results from  increasing  the
rate of the reaction have been demonstrated also through numerical simulation 
 for the reversible reaction $A+B
\leftrightarrow  
A+C$.  We use for that purpose  the two-dimensional concentric billiard 
in which the two modes of
possible reflections represent the two sides of the reaction. It has been
 shown that by repeating either side of the reaction a large 
number $N$
of times (whereas   the other side was repeated with a much lower  rate)  
we obtain, 
in the limit of the largest rate of repetition,  a result 
as if the fastly repeated reaction were not performed at all. \par
The obtained results  conform to  the Zeno effect 
 which
is considered  in \cite{Simonius78}  
to  hold  also in  classical and macroscopic phenomena.  In this effect
(\cite{Misra77}, \cite{Simonius78}, 
   \cite{Pascazio94},  \cite{Itano90})  the
very large number of repeating the same  experiment, in a finite total time, 
  results in 
preserving  the system in the state it was before initiating  these repetitions. 
Moreover, it has 
been shown in  
(\cite{Aharonov80},  \cite{Facchi99}) that this effect can be generalized to 
a whole path of
states   in which 
the final state is different from the initial one. That is, the mechanism of
dense measurement  causes the "realization"  of this specific  path from a very 
large number
of different ones. This result was obtained  by calculating the joint 
probability
of $n$ different reactions,   each of which occurs at its specific place and time, 
where  we
see that densely performing   these reactions 
 causes the joint probability for the occurence of the $n$ 
reactions (where $n \to \infty$) to tend to the
unity value.       
\bigskip \noindent \protect \section*{\bf Acknowledgement }
\bigskip  \noindent   I wish to thank L. P. Horwitz  
 for discussions on this subject and for his review of the manuscript. 
 
 \newpage 

\bigskip \bibliographystyle{}

\end{article}

\end{document}